\documentclass[1p,number,sort]{elsarticle_arXiv}

\usepackage[latin1]{inputenc} 
\usepackage{todonotes}
\usepackage{caption}
\usepackage{lipsum}
\usepackage{float} 

\usepackage{siunitx}

\usepackage{gensymb}

\usepackage{subfig}
\usepackage{graphicx}

\usepackage{amsmath}

\usepackage{mathtools}
\DeclarePairedDelimiter{\evdel}{\langle}{\rangle}

\usepackage[english,breaklinks=true,colorlinks=true]{hyperref} 
\usepackage[hyphenbreaks]{breakurl} 
\usepackage{xcolor}
\definecolor{c1}{rgb}{0,0,1} 
\definecolor{c2}{rgb}{0,0,1} 
\definecolor{c3}{rgb}{0,0,1} 
\hypersetup{
    linkcolor={c1}, 
    citecolor={c2}, 
    urlcolor={c3} 
}
\usepackage{amsmath,amssymb} 
\usepackage{interval} 
\intervalconfig{
soft  open  fences ,
separator  symbol=,,
}

\usepackage[version=3]{mhchem}

\usepackage{longtable}
\usepackage{supertabular}
\usepackage{amssymb}
\usepackage{amsfonts}
\usepackage{parskip}

\renewcommand{\d}{{\,\rm  d}}

\newcommand{\pd}[2]{\displaystyle\frac{\partial #1}{\partial #2}}

\newcommand{\fempty}[1]{{}}

\newcommand{\sty}[1]{\mbox{\boldmath $#1$}}

\newcommand{\fx}{\sty{ x}}

{\begin{equation}\begin{array}{c}}%
{\end{array}\end{equation}}

\usepackage{lineno,hyperref}

\journal{arXiv}

\begin{document}

\begin{frontmatter}

\title{Determining water mass flow control strategies for a turbocharged SI engine using a two-stage calculation method}

\author[mymainaddress]{Peter H\"olz\corref{mycorrespondingauthor}}
\cortext[mycorrespondingauthor]{Corresponding author. Tel.: +49 711 911 88680}
\ead{peter.hoelz@porsche.de}

\author[mysecondaryaddress]{Thomas B\"ohlke}
\author[mymainaddress]{Thomas Kr\"amer}


\address[mymainaddress]{Porsche AG, Porsche Motorsport, Porschestr. 911, 71287 Weissach, Germany}
\address[mysecondaryaddress]{Chair for Continuum Mechanics, Institute of Engineering Mechanics, Karlsruhe Institute of Technology (KIT), Kaiserstr. 10, 76131 Karlsruhe, Germany}

\begin{abstract}
Reduction of heat and friction losses is a proven approach to increase the engine efficiency. Therefore, and due to a stabilized, robust combustion, a specific adjustment of component temperatures is desirable in highly transient conditions.\\
In this paper, a turbocharged SI engine is investigated numerically concerning the potential regulation of temperatures, including heat fluxes, only by controlling the water mass flow rate. Using two independent models, a simplified lumped capacity model and a detailed three-dimensional CFD-CHT simulation, an efficient, two-stage calculation method is suggested for an optimized determination of control strategies and their parameters. This complements existing published works which usually control more than one parameter, but use one model. Different control strategies, like feed forward or feedback controllers, are proposed and compared. In addition, a more holistic approach is presented performing a Monte Carlo simulation which evaluates temperatures, as well as hydraulic pumping losses.\\
Using expedient control strategies and parameters, it could be shown that the engine temperatures can be effectively regulated within a wide range. The two different models show partly similar results, and the efficient, two-stage optimization method has proven its worth. However, there are some significant differences between simplified and detailed modelling which are worth mentioning.
\end{abstract}


\end{frontmatter}

\begin{center}\textnormal{{\scriptsize Copyright, including manuscript, tables, illustrations or other material submitted as part of the manuscript, is assigned to the authors.}}\\[2ex]\end{center}

\tableofcontents

\section{Introduction}
\label{intro}

\subsection{Motivation and state of the art}
\label{State}

In order to increase the engine efficiency by reducing unnecessary heat and friction losses and decrease its emissions by effective exhaust gas treatments under a transient drive, lots of innovative cooling concepts can be found in literature. A reduction of heat losses by application of temperature swings insulation is presented in \cite{Wakisaka2016}: special materials in the combustion chamber with low thermal conductivities and heat capacities decrease heat transfer by reducing the temperature difference between gas and solid during combustion. In \cite{Weise2017}, a concept is presented which uses the phase transition of water to control engine wall temperatures. In addition, the combustion control can be effectively stabilized under dynamic engine conditions with varying load points. This can be explained by the fact that the inlet air temperature, and hence the in-cylinder heat transfer, highly influences the ignition and combustion process. As an example, \cite{Haraldsson2004} investigates closed-loop combustion control mechanisms in a HCCI engine by using thermal management for the inlet air temperature. A review about engine cooling technologies can be found in \cite{Pang2004a}. Different aspects, partially countervailing effects, are discussed. In addition to the already mentioned points, engine protection modes, e.g., improved knock protection, as well as higher volumetric efficiencies for low temperature set points, are discussed. In this context, with regard to unnecessary heat and pumping losses, \cite{Chen2017Structure} investigates numerically optimal cooling structures within an internal combustion engine for given maximum component temperatures. Within the framework of an own developed 1D lumped capacity model, \cite{KangAhnMin2015} also studies two different coolant structures for the engine block and head. \cite{MOHAMED20161352} investigates experimentally the effect of active engine thermal management on a bi-fuel engine. A positive effect on heat release rates, fuel consumption and emissions can be proved. A specific application of thermal management systems, including waste heat recovery, for hybrid electric vehicles with a continuously variable transmission is reported in \cite{Park2013}. In this case, a lumped capacity model is used to increase the transmission efficiency.\\

Classical thermal management strategies for combustion engines try to control the water inlet temperature by installing a thermostat, which can direct a part of the water mass flow around the radiator. In   
\cite{Yang1996Coolant} this concept is supplemented with a pump throttle in order to additionally control the heat transfer coefficients. Analogously, \cite{Wagner2003SAE} investigates a variable speed electric pump by using a reduced order multiple node lumped parameter resistor-capacitor thermal model. In \cite{Setlur2003}, a nonlinear controller for such applications, additionally adjusting the radiator fan speed, is proposed for cases with unmeasurable heat inputs from the combustion process. A global asymptotic regulation of the engine temperature can be shown. Similarly, a robust controller design with an estimate of future disturbances is presented in \cite{Karnik2015}. Using a zero-dimensional model, \cite{Pizzonia2016} presents a mathematically well formulated controller design procedure for the coolant flow rate, including a discretization of the engine map and its piecewise linearization. A one-dimensional, transient numerical model is also developed in \cite{ZhouLan2015}, including some strategies for feed forward and feedback controllers. \\

For these kinds of cooling system, \cite{Haghighat2018} reports a lower fuel consumption of 1.1 percent under NEDC (New European Driving Cycle) cycle operation conditions as well as a reduction of emissions in the range of 5 percent for hydrocarbon and 6 percent for carbon monoxide. In addition, a simplified, holistic vehicle model is proposed, which is mainly based on experimental measurements. Similarly, \cite{Benjac2014Wurzenberger} developes a transient vehicle model with diverse submodels and corresponding interactions. Again, the application of an electrical water pump results in a fuel saving of about 0.75 to 1.1 percent. Additionally, an increase of the turbine outlet temperature can be observed, and hence, a faster catalyst heat-up. A 1D/3D simulation method for the vehicle integrated thermal management is presented in \cite{LuWang2016} and \cite{WangGaoZhang2016}. An effective coupling between 1D submodels of single subsytems with the 3D underhood structure offers suitable boundary conditions and parameters for the simulation. \cite{Caresana2011Num} defines a "perfect cooling system" for vehicles concerning fuel consumption and simulates, with the help of an own developed model, potential savings for different cooling systems.\\

Concerning effective control strategies, there exist many different approaches and the field of control technology is large. The well-known rules after Kessler are based on direct modifications of the compensated system in the frequency space: the so called symmetrical optimum is proposed in \cite{Kessler1958}, whereas \cite{Kessler1955} eliminates the largest time constant in the system. Another, completely empirical, approach is presented by \cite{Ziegler1995}: the research results are based on many test cases with varying parameters. Using a frequency domain model of the plant, a new method for the auto-calibration of PI and PID controllers is presented in \cite{VODA199541}. Some advantages with regard to the above mentioned Ziegler-Nichols approach are reported. Concerning some disadvantages of current design techniques for model predictive control, \cite{KOTHARE1996} proposes a more robust method to deal explicitly with plant model uncertainties. Using dimensional analysis, \cite{FISER2015} developes a quite general  formulation of synthesis of PID control loops with delay. The focus is additionally set on disturbance rejections. A more specific, robust tuning method for first order systems with delay is presented in \cite{YILDIRIM2015}, supplemented by a consideration of stability.

\subsection{Outline of the paper}
\label{Outline}

Concerning control and feedback control systems for engine temperatures, one of the goals of the present paper is the systematic investigation of their controllability by adjusting the water mass flow rate within the water jacket, e.g., at a constant water inlet temperature. Based on that, the dynamic behavior of component temperatures and the internal heat fluxes are studied under transient engine applications. More specifically, typical racing scenarios with their characteristic alternation between braking and acceleration is investigated.  \\
Therefore, contributing to the completeness of already published works, a two-stage calculation method is proposed and explained in detail. In a first step, a simplified lumped capacity model is developed. With the help of the stationary solution and linearization, a first estimation of suitable strategies and parameters for the feed forward and feedback controllers are gained. Afterwards, a Monte Carlo simulation is performed to get optimized control parameters: in a more holistic approach, their influences on engine temperatures, hydraulic flow losses and heat fluxes are investigated. In a second step, the most promising constellations are used for validation and for a more detailed investigation of the aforementioned questions. Therefore, a three-dimensional CFD-CHT engine simulation is used, which has already been validated in \cite{TransientPeterArXivI}. In this reference, using a dynamic engine dynamometer, temperature measurements were conducted with an engine which was equipped with 70 thermocouples. The validation contained highly transient engine conditions, as it is the case in this paper, as well as well-defined stationary boundary conditions. The latter was used for the validation of equation (\ref{SeparationApproach}). Hence, because of the two completely independent models, the two-stage method can be seen as a kind of validation for the simplified, lumped model. \\
The research question can be formulated as follows: With regard to engine temperatures near the combustion chamber, which temperature range, including the temperature deviation, its dynamic behavior and the resulting heat flows, is feasible only by adjusting the water mass flow rate at a constant water inlet temperature? On this basis, what are the differences between the fast lumped capacity model, which is suitable for optimization and sensitivity calculations, and a more compute-intensive, three-dimensional CFD-CHT simulation?

\section{Optimized thermal management due to various feed forward and control strategies}
\label{Thermo}
\setlength{\parindent}{0pt}
\setlength{\parskip}{0pt}	

\subsection{Method used in this paper}
\label{Method}
\setlength{\parindent}{0pt}
\setlength{\parskip}{0pt}	

\begin{figure}[H]
\captionsetup{width=1.0\textwidth}
\begin{center}
\includegraphics[width=1.0\textwidth]{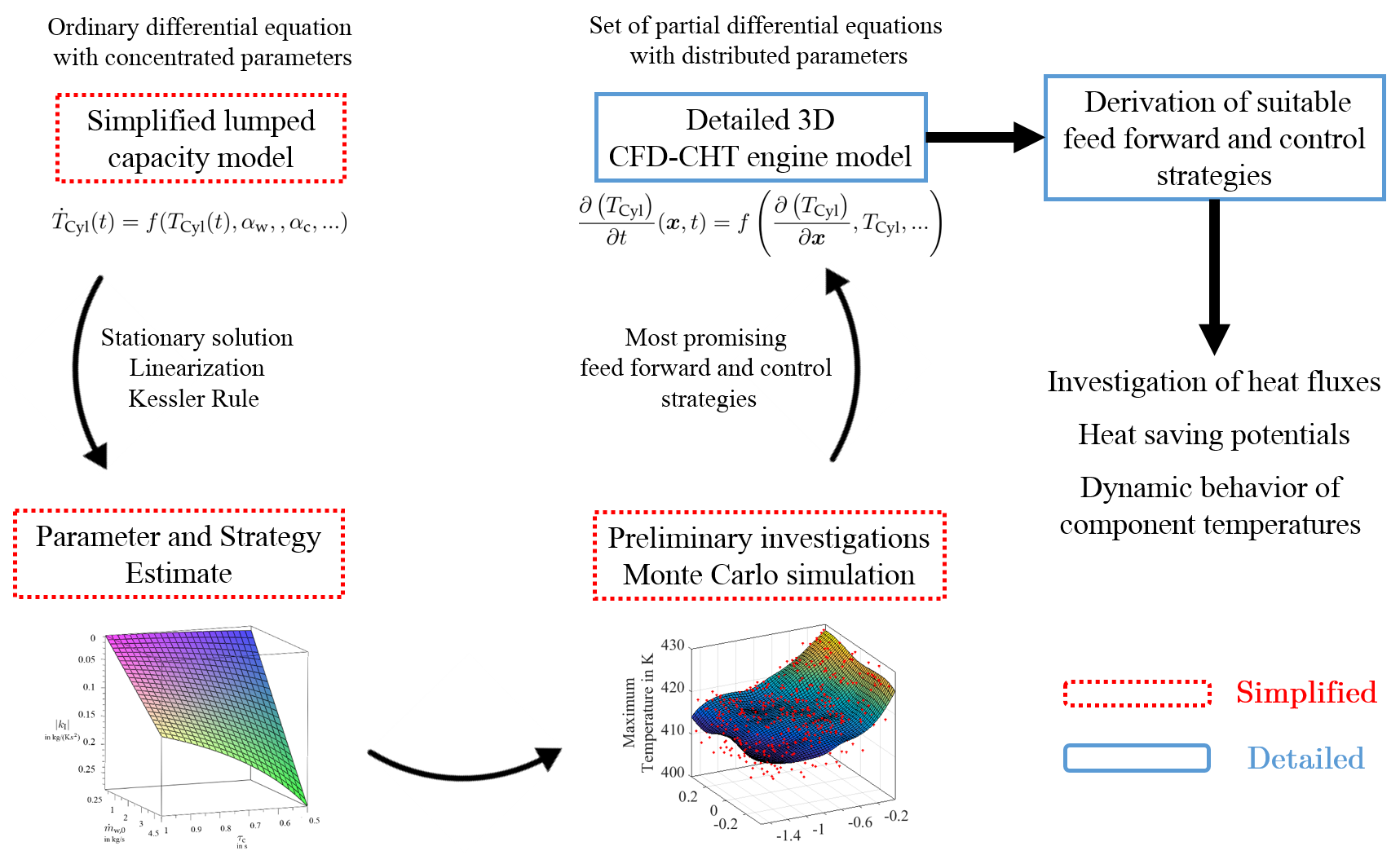} 
\end{center}
\caption{Proposed two-stage calculation method for an optimized thermal management.}
\label{fig:Temperature_Method}
\end{figure}

Various feed forward and control strategies for the engine water mass flow rate are investigated numerically and compared with a mechanical water pump with a fixed transmission ratio. Fig. \ref{fig:Temperature_Method} shows the proposed method. A simplified lumped capacity model of the engine serves as a first parameter and strategy estimation with subsequent, preliminary investigations, including a Monte Carlo simulation. Afterwards, the most promising parameter sets and strategies are investigated with detailed CFD-CHT engine simulations. Special focus is laid on heat saving potentials and the dynamic behavior of diverse component temperatures under a transient drive.

Therefore, simplified and the detailed, three-dimensional simulations were carried out for one representative race lap and analysed regarding various criteria such as heat losses, solid temperature fluctuations or the necessary hydraulic power. 

\subsection{Simplified thermal model - lumped capacity model}
\label{ThermalModel}

The used heat transfer models are extracted from \cite{PeterSimilar2017} and \cite{TransientPeterArXivI}. Consequently, the heat transfer coefficient for the water jacket can be written as:

\begin{equation}
\alpha_{\text{w}}(\fx,t) \approx \alpha_{\text{ref}}(\fx) \left(\frac{\dot{m}_{\text{w}}(t)}{\dot{m}_{\text{w}}|_{\text{ref}}}\right)^m.
\label{SeparationApproach}
\end{equation}

This approach separates the time and location dependency by means of a turbulence Reynolds exponent $m=0.7$. The subscripts describe representative reference states. The location dependency is omitted because of the lumped-capacity model and, therefore, serves as a calibration parameter with a certain physical meaning: choosing $\dot{m}_{\text{w}}|_{\text{ref}}$ as a representative mean water mass flow rate, a range of $\alpha_{\text{ref}}\sim~10^4$ W/(m\textsuperscript{2}K) is used. In this case, the HTC is referred to the water temperature as the reference temperature. \\
Following approach is used for the combustion chamber HTC:

\begin{align}
\langle \alpha_{\text{c}}  \rangle (t) = \int\limits_{\mathbb{R}_{\geq0}}  A p_{\alpha|n(t)}(A)  \,\d A.
\label{TransientIntegrationComplete}
\end{align}

During transient simulations, the conditional probability density function for the heat transfer coefficient $p_{\alpha|n(t)}$ has to be used. It strongly depends on the engine speed $n(t)$ and the engine load. However, due to the fact that a race engine is investigated, which spends most of the time in a complete full load or coasting condition, part load conditions are neglected. For the investigated race track its percentage was less than 10 percent. Because of the cyclic fluctuations in the pressure curves, the HTC is interpretated as a random variable with its realisation $A$. For the reference temperature, in this case the gas temperature $T_{\text{gas}}$ within the combustion chamber, an analogous procedure was done. However, due to the nonlinear nature of heat transfer, a statistically modified reference temperature $T_{\text{mod}}=\evdel{\alpha_{\text{c}} T_{\text{gas}} } / \evdel{\alpha_{\text{c}}} $ has to be used. In the following, it is called ACT (Average Cylinder Temperature). In this paper, the determination of $p_{\alpha|n(t)}$ is based on the Woschni model \citep{Woschni1967} and high pressure indication measurements. Details can be found in \cite{PeterSimilar2017} and \cite{TransientPeterArXivI}. Coasting conditions are treated in a similar manner.

During simulation, following ordinary differential equation for the engine temperature $T_{\text{Cyl}}$ is solved numerically:

\begin{align}
C_{\text{v}} \rho \Delta x \dot{T}_{\text{Cyl}}(t)& =\chi  \evdel{\alpha_{\text{c}}}  (t) \left( T_{\text{mod}}(t) - T_{\text{Cyl}}(t) \right) + \alpha_{\text{w}} (t)   \left( T_{\text{w}}(t) - T_{\text{Cyl}}(t) \right),\label{DGL_MOdelLumped}\\
T_{\text{w}}(t)& =T_{\text{w,i}}+\dot{Q}(t)/\left( \dot{m}_{\text{w}}(t) C_{\text{p,w}} \right).
\label{TW_Approx}
\end{align}

\begin{figure}[H]
\captionsetup{width=1.0\textwidth}
\begin{center}
	  \includegraphics[width=0.8\textwidth]{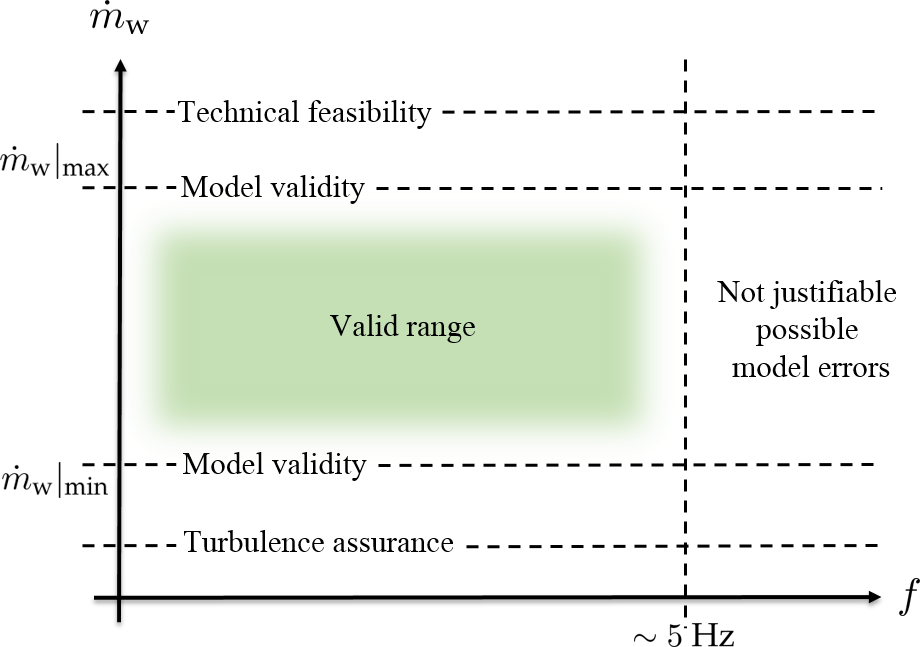} 
\end{center}
\caption{Allowable range for the water mass flow rate during the simulation in the time ($\dot{m}_\text{w}|_{\text{min}}$ and $\dot{m}_\text{w}|_{\text{max}}$) and frequency ($5$ Hz) domain.}
\label{fig:Parameter_Space_Water}
\end{figure}

$T_{\text{w,i}}$ is the inlet water temperature, and serves as a model input parameter. $\dot{Q}$ is the heat flux which is transferred from the solid to the water. 
The time-dependent variables $\alpha_{\text{c}}$ and $\alpha_{\text{w}}$ are the effective heat transfer coefficients according to equation (\ref{TransientIntegrationComplete}) and (\ref{SeparationApproach}). The parameter $\chi$, which should be smaller than one, describes the ratio between the effective surfaces of the combustion chamber and the water jacket. It serves as an additional calibration parameter. In this case, a value of 0.3 was chosen. $\Delta x$ is the characteristic wall thickness, and is necessary for dimensional reasons. It can be seen as the volume of the representative engine part, which is described by a lumped capacity, divided through the effective surface area of the water wetted side. Therefore, it is not completely independent, e.g., it has to be scaled according to the existing geometry. In this paper, a value of $\Delta x=15$ mm is used. $C_{\text{v}}$ and $\rho$ are the heat capacity and the density of the solid, respectively. Typical values for aluminium cylinder heads are in the range of 900 J/(kgK) and 2.7 kg/m\textsuperscript{3}.\\
There are lots of system parameters which influence heat flow: Prandtl and Reynolds number, oscillating frequency and amplitude, as well as the entry length of pipes. Experimental results with comparable parameters prevailing in engine water channels can be found in \cite{Patel2016} or \cite{Habib2004}. Choosing $Pr \approx 1$, $Re \approx 8~10^3-5~10^4$ and the oscillating frequency in the range of $1-5$ Hz, the Nusselt number increase, respectively decrease, can be about 15 percent. In order to ensure the model validity and the technical feasibility, the water pump is modelled as a PT1-element with limited water mass flow rates. Local Reynolds numbers should be noticeably larger than $2300$, the transition number between laminar and turbulent pipe flow. 
In addition, if the water mass flow rate is very low, the amplitudes of the water temperature will be quite large: see equation (\ref{TW_Approx}). Normally, the water pressure is limited and, therefore, to avoid local boiling and corresponding cavitation damage, a minimum flow rate must be guaranteed. On top of that, the design of a suitable vehicle radiator is much complicated for large oscillations, and fluctuating water inlet temperatures are undesirable for real engines. The transfer function for the water pump in Laplace space is 

\begin{equation}
G_{\text{p}}=\frac{1}{\tau_{\text{p}} s+1},
\label{TransferWAPU}
\end{equation}

with a characteristic time constant $\tau_{\text{p}}$, which was set to $0.2$ s, resulting in a base frequency of $5$ Hz: the allowable range for the water mass flow rate during the simulation is qualitatively shown in fig. \ref{fig:Parameter_Space_Water}. 

\subsection{Detailed thermal model - CFD-CHT method}
\label{Method}
\setlength{\parindent}{0pt}
\setlength{\parskip}{0pt}
The detailed finite volume model of the engine consists of all relevant engine components like the crank case, cylinder heads, valves, pistons, as well as the cylinder liners. Details about the model can be read in \cite{TransientPeterArXivI}. The transient, thermal boundary conditions in the combustion chambers are given according to equation (\ref{TransientIntegrationComplete}) with some spatial extensions. Details can be found in \cite{PeterSimilar2017}. In addition, one can find the used meshing strategy, including local mesh refinements, and a mesh study for the cylinder head. The thermal boundary conditions for the engine water jacket are given by the SST $k$-$\omega$ turbulence model by Menter \citep{Menter1994}. \\
The adaption of the water reference temperature is also modelled according to equation (\ref{TW_Approx}). The operating point of the water pump is also limited according to fig. \ref{fig:Parameter_Space_Water}.  Therefore, the transfer function (\ref{TransferWAPU}) is used. 

\subsection{Feed forward strategy}
\label{Forward}

\begin{figure}[H]
\captionsetup{width=1.0\textwidth}
\begin{center}
	  \includegraphics[width=0.9\textwidth]{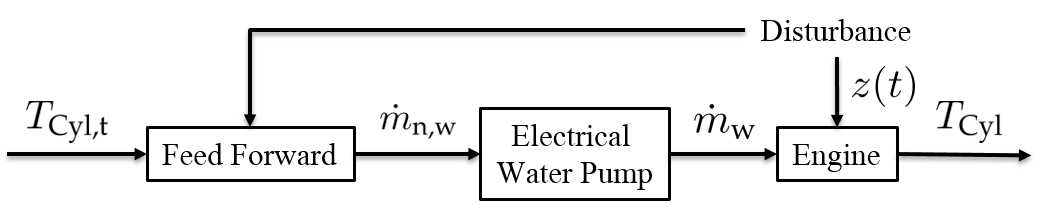} 
\end{center}
\caption{Block diagram for the feed forward strategy.}
\label{fig:Forward_Diagram}
\end{figure}

The block diagram for the feed forward strategy is shown in fig. \ref{fig:Forward_Diagram}. 
The time-dependent disturbance $z(t)$ on the engine are transient heat fluxes caused by alternating firing and coasting conditions. The control variable is the cylinder head temperature $T_{\text{Cyl}}$ of a representative measuring position close to the combustion chamber. 
The correcting variable for all scenarios is the water mass flow rate $\dot{m}_{\text{w}}$. $T_{\text{Cyl,t}}$ is the command variable. $\dot{m}_{\text{n,w}}$ is the nominal water mass flow rate.

Neglecting changes in the water temperature according to equation (\ref{TW_Approx}), and setting $\dot{T}_{\text{Cyl}}=0$ and $T_{\text{Cyl}}=T_{\text{Cyl,t}}$ in equation (\ref{DGL_MOdelLumped}), it follows with equation (\ref{SeparationApproach}) for the nominal water mass flow rate 

\begin{equation}
\dot{m}_{\text{n,w}}(t)=\left(\frac{\chi  \evdel{\alpha_{\text{c}}}  (t) \left( T_{\text{mod}}(t) - T_{\text{Cyl,t}} \right)}{ -  \left( \alpha_{\text{ref}}/\dot{m}_{\text{w}}|_{\text{ref}}^m \right)  \left( T_{\text{w,i}} - T_{\text{Cyl,t}} \right)}\right)^{1/m}.
\label{AuxWaterMAss}
\end{equation}

Due to fig. \ref{fig:Parameter_Space_Water}, there are some restrictions regarding the feasible mass flow. 

\subsection{Control strategy}
\label{Controlling}

\begin{figure}[H]
\captionsetup{width=1.0\textwidth}
\begin{center}
	  \includegraphics[width=1.0\textwidth]{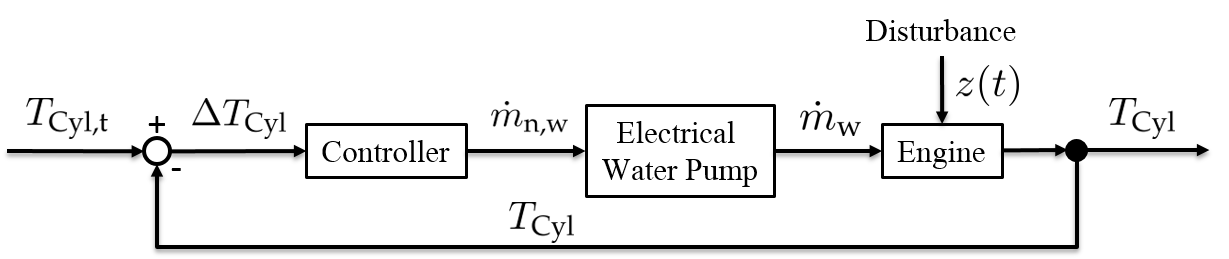} 
\end{center}
\caption{Block diagram for the closed control loop.}
\label{fig:Control_Diagram}
\end{figure}

The block diagram for the closed control loop is shown in fig. \ref{fig:Control_Diagram}. It is assumed that there are no measurement errors or delay elements in the 
feedback path. 

The controller is executed as PID control device with following transfer function in Laplace space:

\begin{equation}
G_{\text{c}}=k_{\text{P}}+\frac{k_{\text{I}}}{s}+k_{\text{D}}s,
\label{Controler_Basic}
\end{equation}

with $k_{\text{P}}$, $k_{\text{I}}$ and $k_{\text{D}}$ as the control parameter for the various components of the controller. Equation (\ref{Controler_Basic}) can be rewritten in terms of two time constants $T_{\text{R,1}}$ and $T_{\text{R,2}}$:

\begin{equation}
G_{\text{c}}=k_{\text{I}}\frac{(1+T_{\text{R,1}}s)(1+T_{\text{R,2}}s)}{s},
\label{Controler_Basic_Mod}
\end{equation}

with $k_{\text{P}}=k_{\text{I}}(T_{\text{R,1}}+T_{\text{R,2}})$ and $k_{\text{D}}=k_{\text{I}}T_{\text{R,1}}T_{\text{R,2}}$. When choosing suitable control parameters, it must be noted that equation (\ref{DGL_MOdelLumped}) is non-linear in the water mass flow rate $\dot{m}_{\text{w}}$. Therefore, in order to model the engine as a PT1-element with a characteristic time constant $\tau_{\text{e}}$ and a static amplification $k_{\text{s,e}}$, a linearization at a representative operating point is necessary. Neglecting changes in the water reference temperature according to equation (\ref{TW_Approx}), it follows for the engine temperature difference $\Delta T_{\text{Cyl}}$

\begin{align}
\underbrace{     \frac{ C_{\text{v}} \rho \Delta x }{ \alpha_{\text{w,0}} }    }_{=\tau_{\text{e}}} \Delta \dot{T}_{\text{Cyl}}+\Delta T_{\text{Cyl}}&=\frac{ T_{\text{w,0}}-T_{\text{Cyl,0}} }{ \alpha_{\text{w,0}} } \Delta \alpha_{\text{w}}, \label{LinearI} \\
&=\underbrace{  \frac{ T_{\text{w,0}}-T_{\text{Cyl,0}} }{ \alpha_{\text{w,0}} } 0.7 \dot{m}^{-0.3}_{\text{w,0}} \frac{\alpha_{\text{ref}}}{\dot{m}_{\text{w}}|_{\text{ref}}}   }_{=k_{\text{s,e}}} \Delta \dot{m}_{\text{w}}, \label{LinearII}
\end{align}

with $\Delta T_{\text{Cyl}}=T_{\text{Cyl}}-T_{\text{Cyl,0}}$. The subscript 0 corresponds to the representative operating point. The last step is the result of equation (\ref{SeparationApproach}). As a first approximation for the control parameters, the rule according to Kessler \citep{Kessler1955} was used. 

\begin{figure}[H]
\captionsetup{width=1.0\textwidth}
\begin{center}
\subfloat[]{\includegraphics[width=0.5\textwidth]{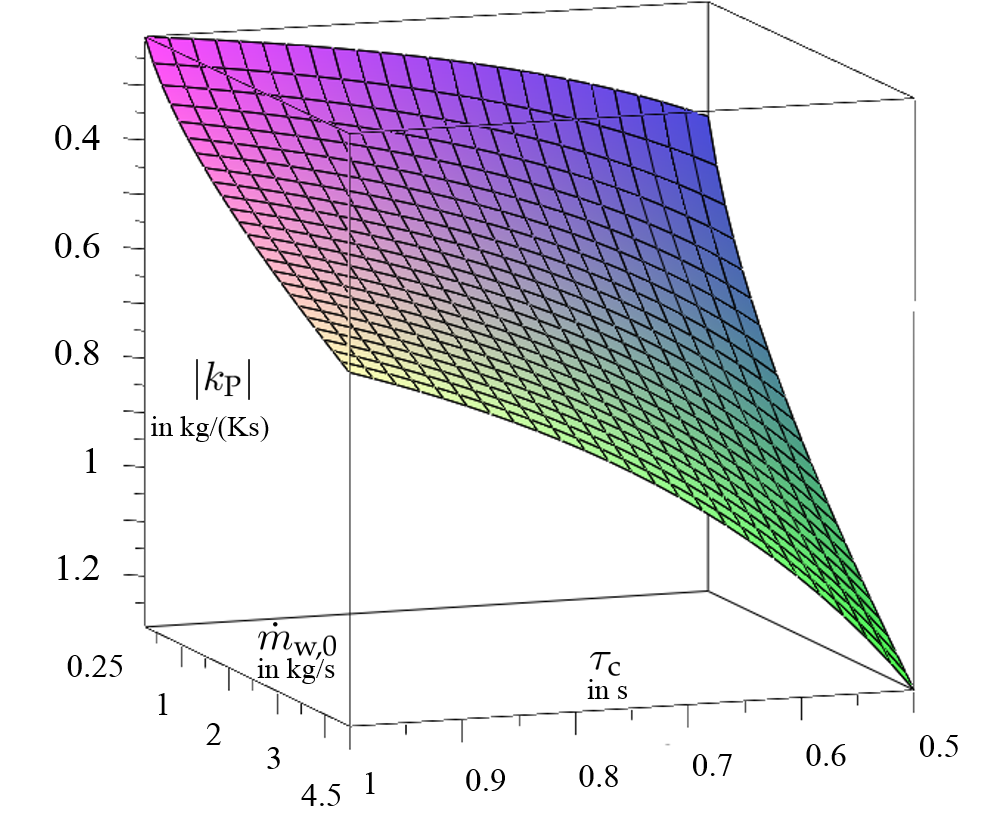}} 
\subfloat[]{\includegraphics[width=0.5\textwidth]{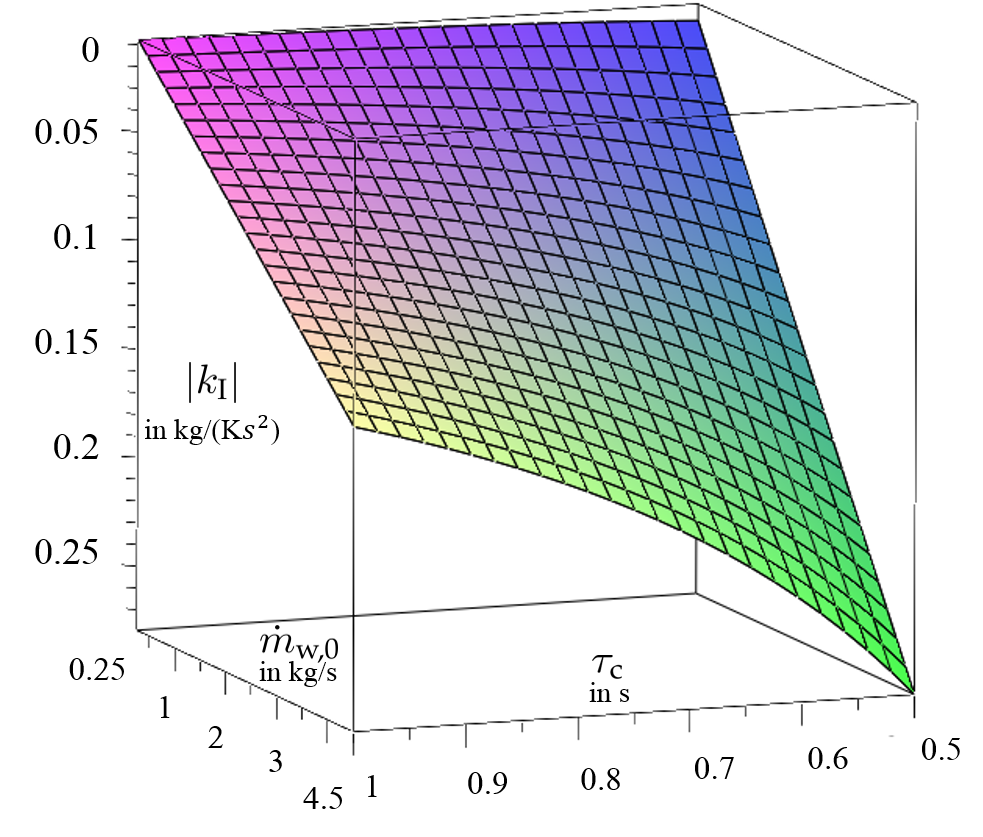}} \\
\subfloat[]{\includegraphics[width=0.5\textwidth]{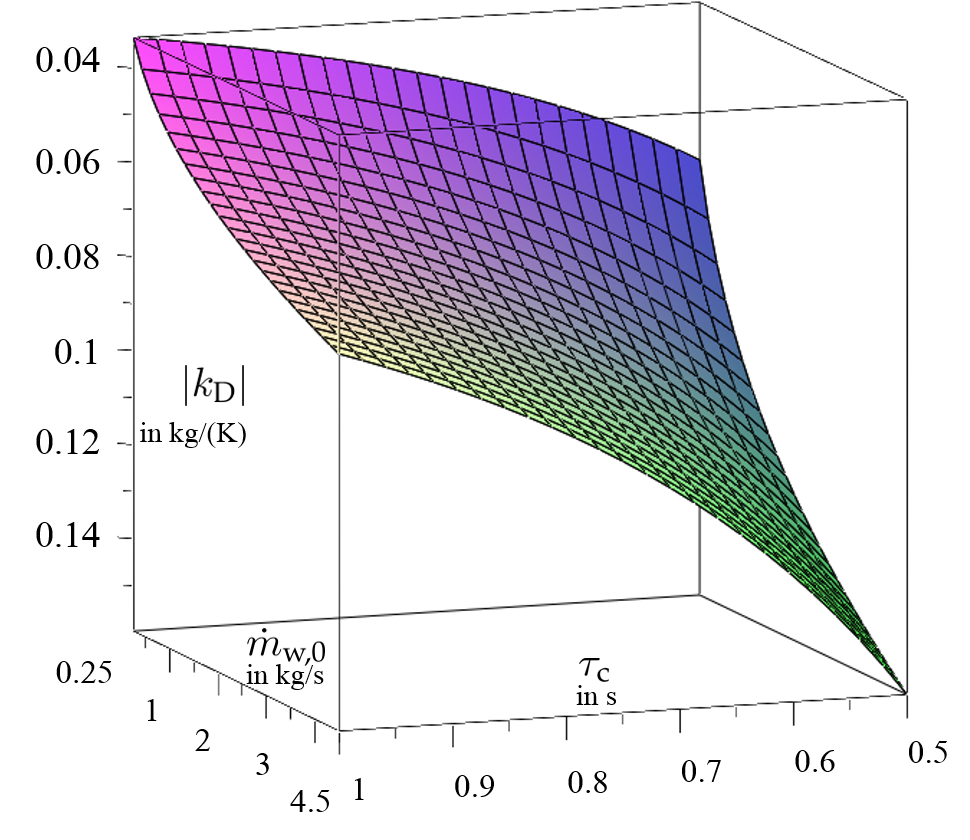}}
\end{center}
\caption{First approximations for the control parameters of the PID control device as functions of  the water mass flow rate in the representative operating point $\dot{m}_{\text{w,0}}$ and the resulting system time constant of the closed control loop $\tau_{\text{c}}=1/(k_{\text{I}}k_{\text{s,e}})$.}
\label{fig:EstimationParaControl}
\end{figure}

Compensating both the time constants of the water pump $\tau_{\text{p}}$ and the engine $\tau_{\text{e}}$, the transfer function for the open control loop $G_{\text{o}}$ and the closed loop $G_{\text{cl}}$ in Laplace space is 

\begin{align}
G_{\text{o}}&=\frac{ k_{\text{I}} k_{\text{s,e}} } { s },\\
G_{\text{cl}}&=\frac{ G_{\text{o}} } { 1+G_{\text{o}} }=\frac{1} {1+\frac{1}{k_{\text{I}}k_{\text{s,e}}}s}.
\end{align}
  
As an example, fig. \ref{fig:EstimationParaControl} shows the three control parameters of the PID control device for a representative engine operating point. The given values should be regarded as first, possible approximations, because of the strong non-linearity and the ever changing engine states. According to equation (\ref{LinearI}) and (\ref{LinearII}), representative  temperatures $T_{\text{w,0}}=373$ K and $T_{\text{Cyl,0}}=419$ K are used for the time constant $\tau_{\text{e}}$ and the amplification $k_{\text{s,e}}$. The control parameters are given as functions of the water mass flow rate in the representative operating point $\dot{m}_{\text{w,0}}$ and the resulting system time constant of the closed control loop $\tau_{\text{c}}=1/(k_{\text{I}}k_{\text{s,e}})$.

\subsection{Feed forward and control strategy}
\label{FeedControlling}

\begin{figure}[H]
\captionsetup{width=1.0\textwidth}
\begin{center}
	  \includegraphics[width=0.9\textwidth]{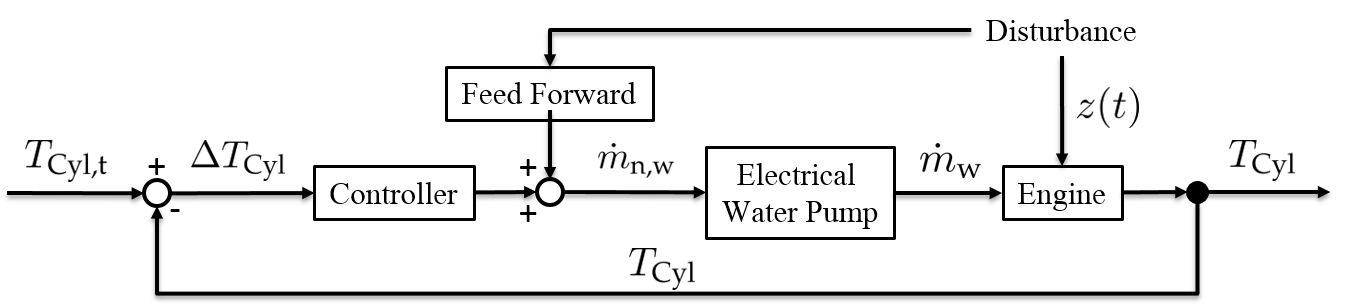} 
\end{center}
\caption{Block diagram for the feed forward strategy including a closed control loop.}
\label{fig:ForwardControll_Diagram}
\end{figure}

Of course, a combination of the aforementioned strategies is possible. Fig. \ref{fig:ForwardControll_Diagram} shows the resulting block diagram.

\section{Results and discussion}
\label{Experiment}
\setlength{\parindent}{0pt}
\setlength{\parskip}{0pt}	

\subsection{Feasible temperature range and consistency}
\label{Feasible}

\begin{figure}[H]
\captionsetup{width=1.0\textwidth}
\begin{center}
	 a) \includegraphics[width=0.45\textwidth]{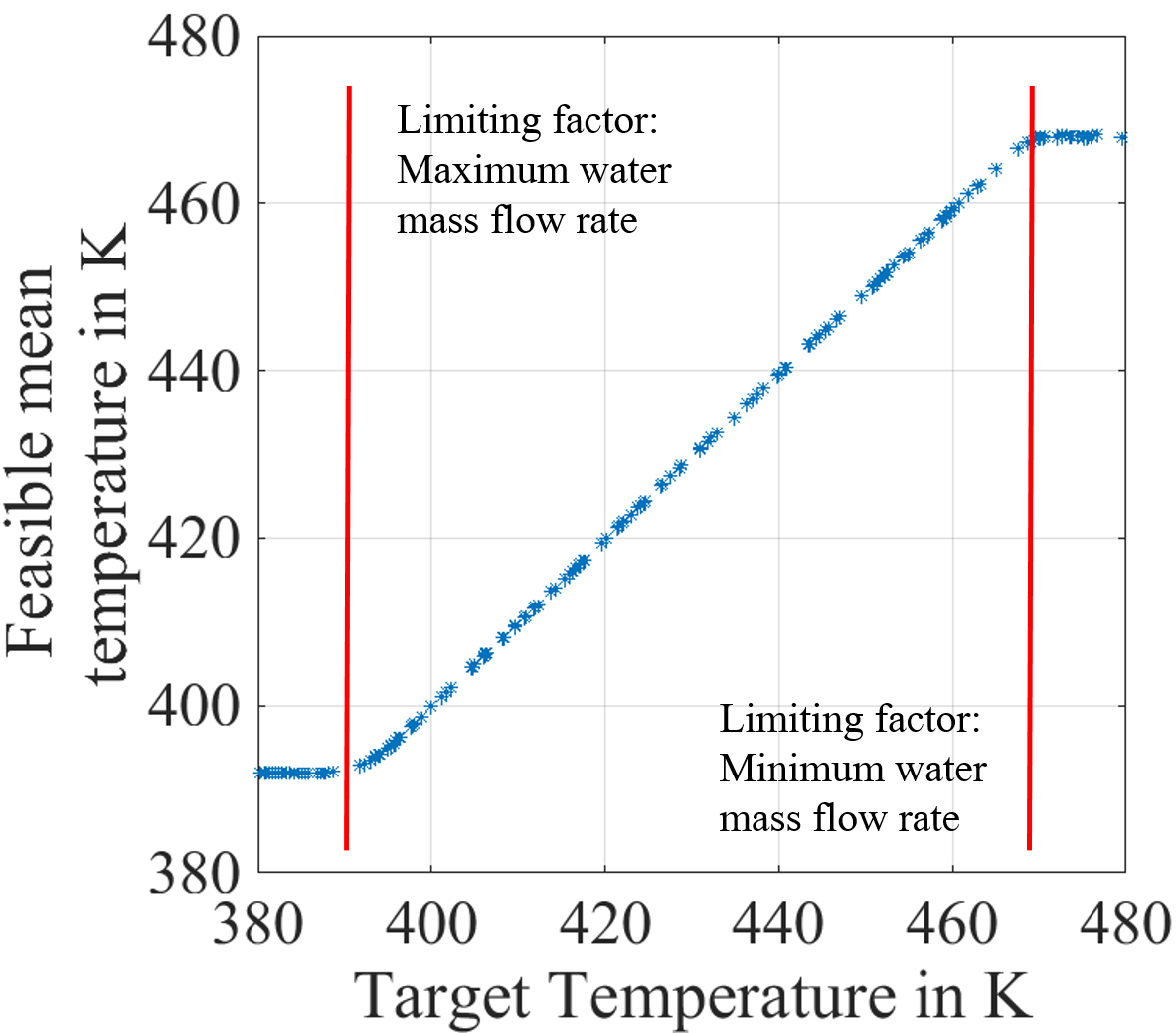} 
     b) \includegraphics[width=0.45\textwidth]{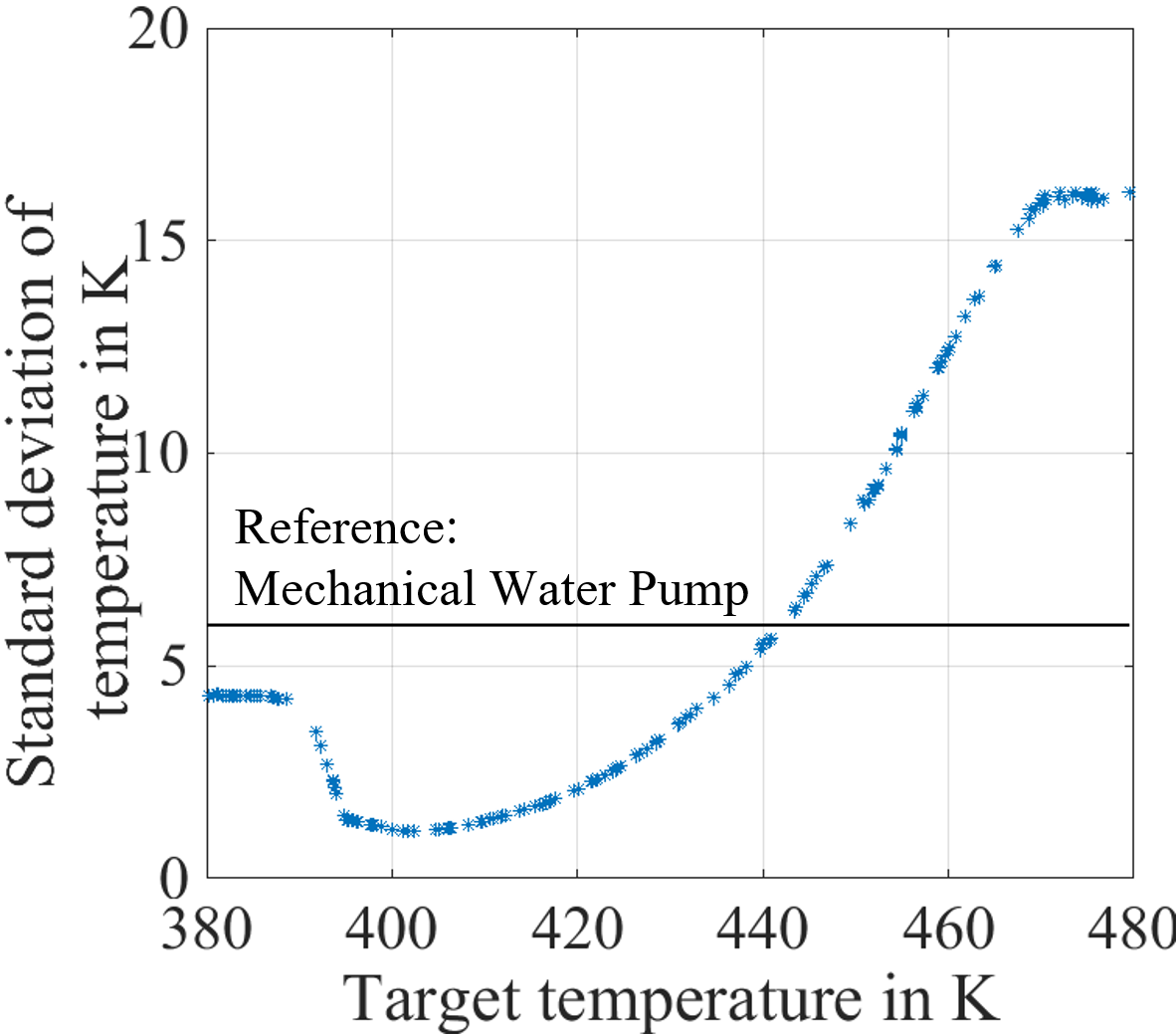}	 
	 c) \includegraphics[width=1\textwidth]{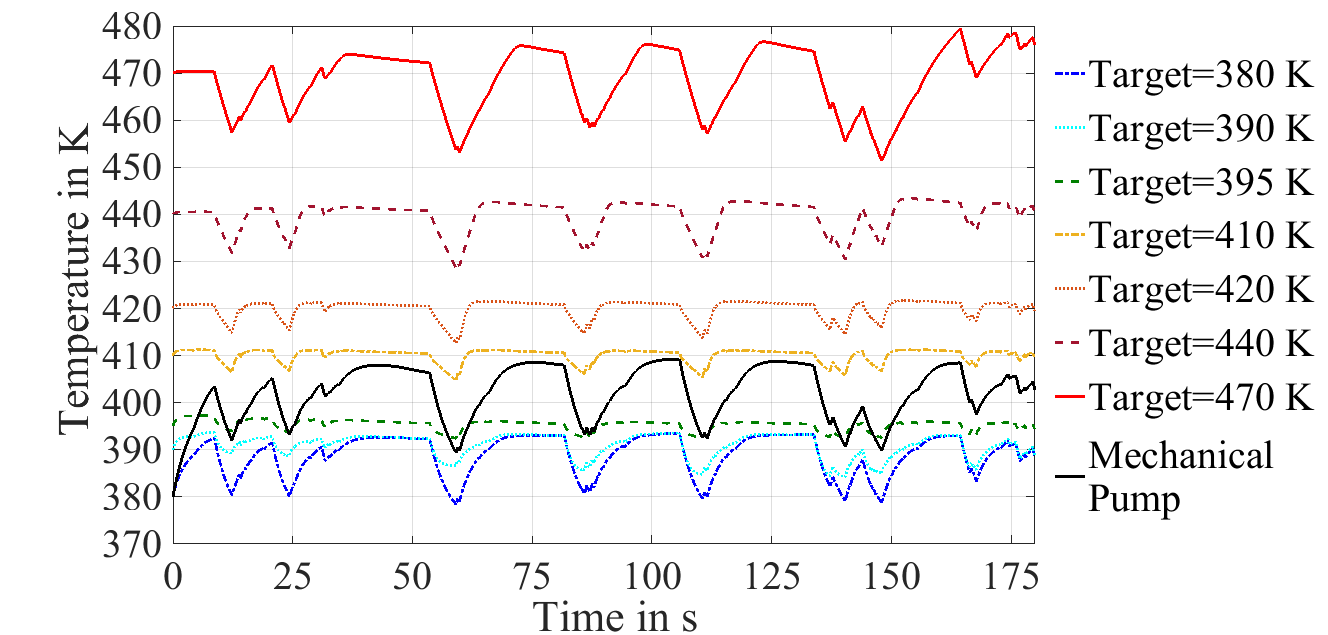}
\end{center}
\caption{Lumped capacity model: feasible temperature range by using a PID controller with $k_{\text{P}}=-1.4~\text{kg/(Ks)}$, $k_{\text{I}}=-0.05~\text{kg/(Ks\textsuperscript{2})}$ and $k_{\text{D}}=-1.0~\text{kg/K}$. a) Mean temperature. b) Standard deviation of the temperature. c) Exemplary temperature curves. The ratio of the (reference) mechanical water pump was set to reach a maximum temperature of 407 K.}
\label{fig:FeasibleRange}
\end{figure}

Firstly, the feasible temperature range and the temporal behaviour was investigated by means of the lumped capacity model. Therefore, the minimum water mass flow rate was set to 0.25 kg/s, whereas the maximum rate was 4.5 kg/s. As can be seen in fig. \ref{fig:FeasibleRange} a), a temperature range about 80 K can be realized by using a water pump with a variable water mass flow rate. The limitations result from the allowable parameter space in fig. \ref{fig:Parameter_Space_Water}.  

However, as can be seen in fig. \ref{fig:FeasibleRange} b), the  standard deviation of the temperature rises strongly with higher target temperatures. Allowing a maximum value of 6 K, which can also be reached by a conventional mechanical water pump, a range of 60 K can be realized. The reasons are the same as for fig \ref{fig:FeasibleRange} a): Targeting a component temperature of 380 K, the resulting water mass flow rate was constant at 4.5 kg/s. One gets exactly the opposite result for a target temperature of 470 K, e.g., a nearly constant mass flow rate of 0.25 kg/s. However, the temporary, local temperature minima are deeper because of larger thermal differences between firing and coasting engine conditions: fig. \ref{fig:FeasibleRange} c) shows some exemplary temperature curves for various target temperatures.

\subsection{Monte Carlo simulation}
\label{MonteCarlo}

\begin{figure}[H]
\captionsetup{width=1.0\textwidth}
\begin{center}
	 \includegraphics[width=1.0\textwidth]{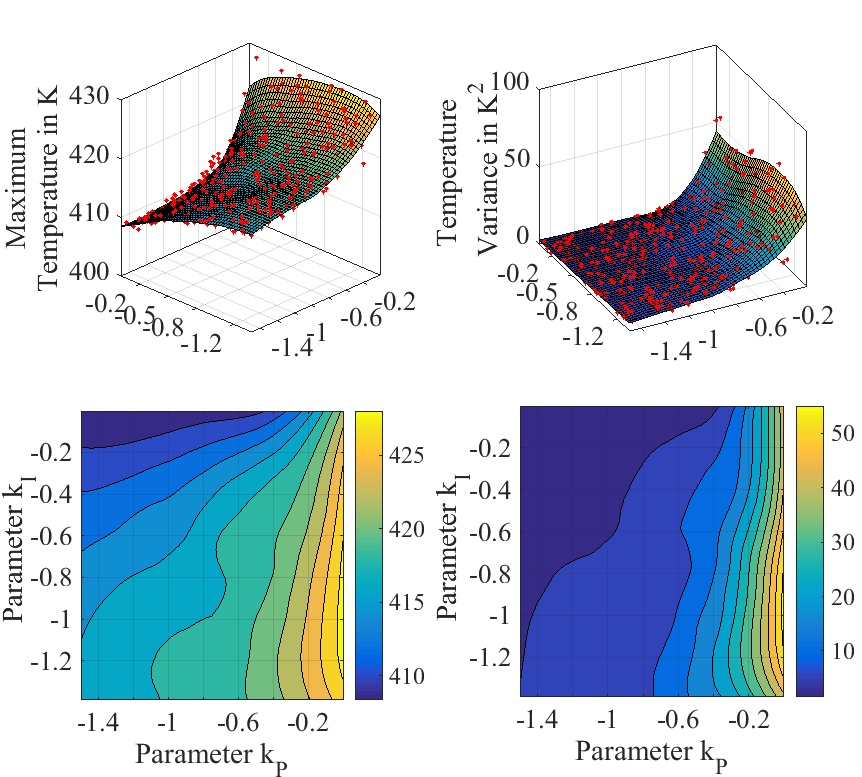} 
\end{center}
\caption{Lumped capacity model: Monte Carlo simulation for a target temperature of 407 K in the $k_{\text{P}}$ - $k_{\text{I}}$ space. The resulting cylinder head temperature is shown.}
\label{fig:MOnteI}
\end{figure}

\begin{figure}[H]
\captionsetup{width=1.0\textwidth}
\begin{center}
	 \includegraphics[width=1.0\textwidth]{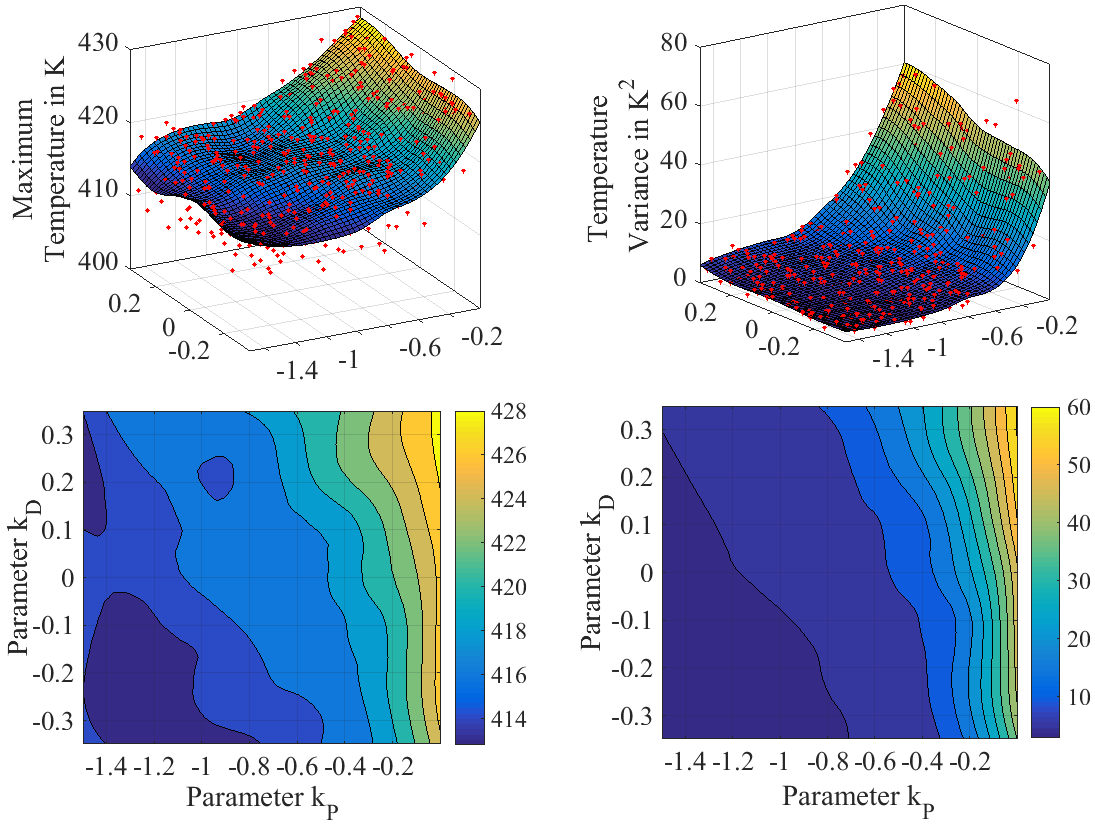}
\end{center}
\caption{Lumped capacity model: Monte Carlo simulation for a target temperature of 407 K in the $k_{\text{P}}$ - $k_{\text{D}}$ space. The resulting cylinder head temperature is shown.}
\label{fig:MOnteII}
\end{figure}

\begin{figure}[H]
\captionsetup{width=1.0\textwidth}
\begin{center}
	 \includegraphics[width=1.0\textwidth]{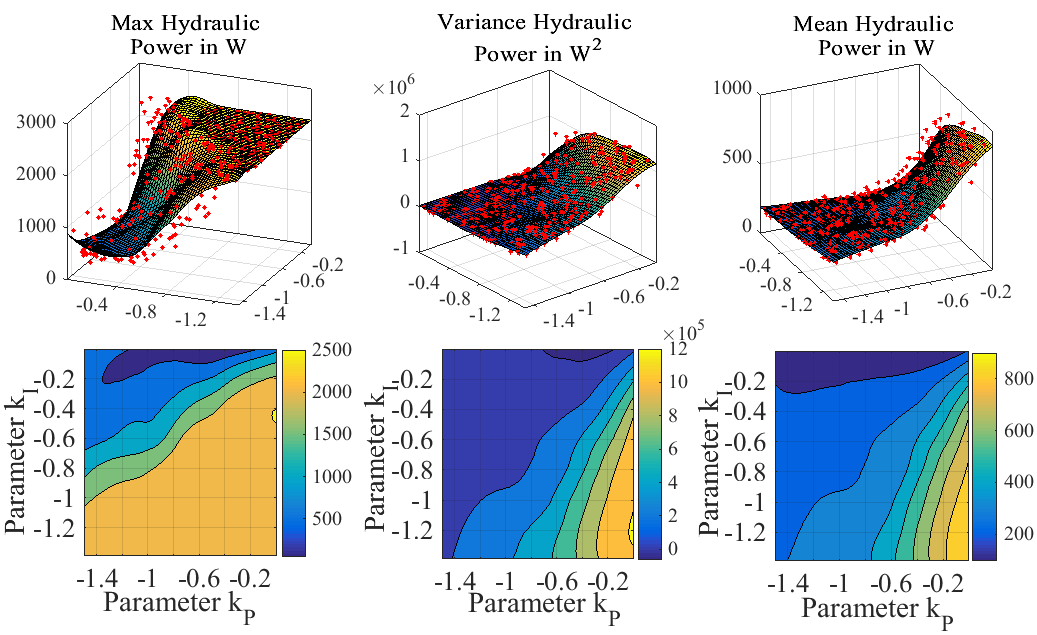} 
\end{center}
\caption{Lumped capacity model: Monte Carlo simulation for a target temperature of 407 K in the $k_{\text{P}}$ - $k_{\text{I}}$ space. The resulting hydraulic power is shown.}
\label{fig:MOnteIII}
\end{figure}

\begin{figure}[H]
\captionsetup{width=1.0\textwidth}
\begin{center}
	 \includegraphics[width=1.0\textwidth]{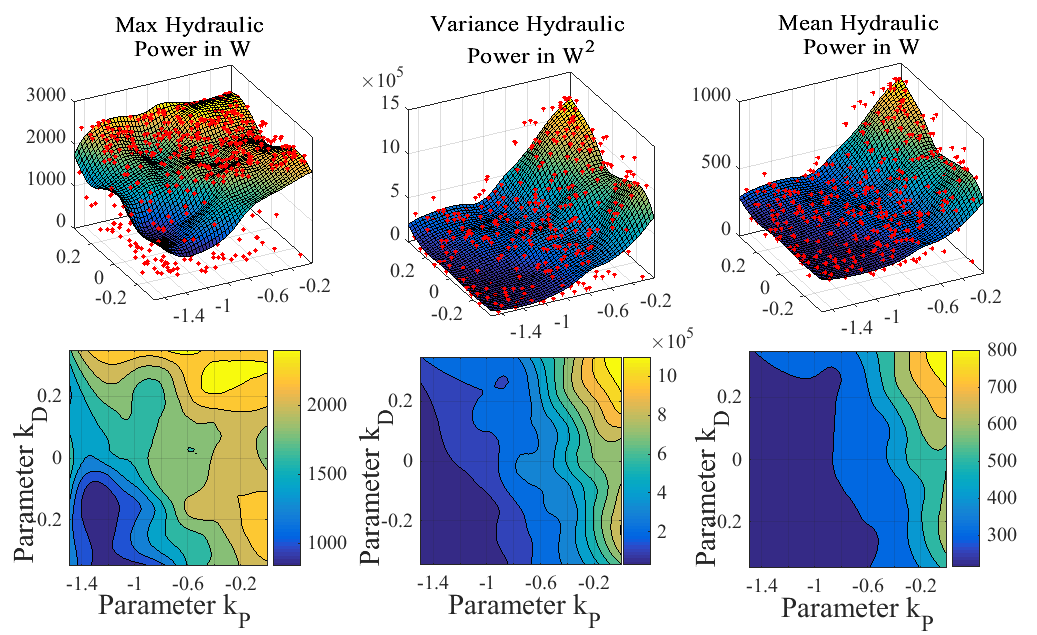} 
\end{center}
\caption{Lumped capacity model: Monte Carlo simulation for a target temperature of 407 K in the $k_{\text{P}}$ - $k_{\text{D}}$ space. The resulting hydraulic power is shown.}
\label{fig:MOnteIV}
\end{figure}

Fig. \ref{fig:MOnteI} to \ref{fig:MOnteIV} show some results of a Monte Carlo simulation for a constant target temperature of 407 K. In order to investigate the sensitivity of the controller parameters according to fig. \ref{fig:Control_Diagram}, the corresponding values of $k_{\text{P}}$, $k_{\text{I}}$ and $k_{\text{D}}$ were varied systematically. The results in fig. \ref{fig:EstimationParaControl} serve as starting values.

Regarding regular temperature curves under a transient drive, high $k_{\text{P}}$ and small $k_{\text{I}}$ absolute values are expedient, see fig. \ref{fig:MOnteI}. However, $k_{\text{P}}$ values higher than 0.6 do not provide a significant advantage if the $k_{\text{I}}$ value is small enough. Because open loops without integrating parts do not lead to stationary accurate closed loops, $k_{\text{I}}$ levels close to zero are not recommended: the limit theorem for Laplace-Transformations

\begin{equation}
\lim\limits_{t \rightarrow \infty}{y(t)}=\lim\limits_{s \rightarrow 0}{s Y(s)},
\end{equation}

for a function $y$ and its Laplace transform $Y$, requires following condition for stationary accuracy: 

\begin{equation}
G_{\text{cl}}(0)=\frac{ G_{\text{o}}(0) } { 1+G_{\text{o}}(0) }=1.
\end{equation}

This can be only fulfilled with $\lim\limits_{s \rightarrow 0}{G_{\text{o}}(s)}\rightarrow \infty$. Therefore, $k_{\text{I}}$ values different to zero are necessary. Values about -0.05 and -0.1 are recommended. In the $k_{\text{P}}$ - $k_{\text{D}}$ space, a similar behavior can be observed: high $k_{\text{P}}$ and $k_{\text{D}}$ absolute values are expedient. With values higher than 1.2 and 0.3, no significant advantages can be observed, see fig. \ref{fig:MOnteII}. \\
Regarding hydraulic power losses, a similar result can be observed in fig. \ref{fig:MOnteIII} and \ref{fig:MOnteIV}. Concerning a holistic consideration, e.g., taking into account the maximum hydraulic power, its variance and mean value, a similar range in the in the $k_{\text{P}}$ - $k_{\text{I}}$ - $k_{\text{D}}$ space is recommended.

\subsection{Dynamic effects}
\label{TransientEffect}

\begin{figure}[H]
\captionsetup{width=1.0\textwidth}
\begin{center}
	 \includegraphics[width=1.0\textwidth]{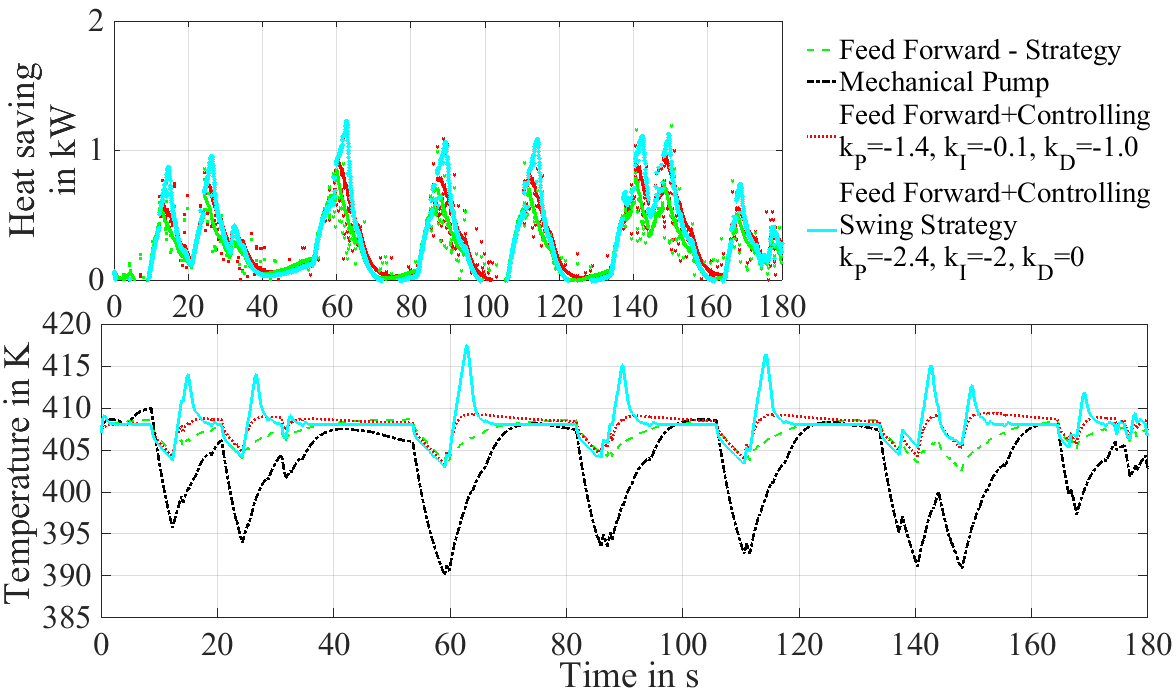} 
\end{center}
\caption{Lumped capacity model: comparison of four different feed forward and control strategies concerning heat saving potentials and temperature consistency under a transient drive.}
\label{fig:DynamicI}
\end{figure}

On the comparability of the following strategies, the boundary condition for all cases is the same: a fixed maximum engine temperature under full load state at the end of straight track. As can be seen in fig. \ref{fig:DynamicI}, the pure feed forward strategy decreases significantly the temperature standard deviation. Especially in coasting situations or in the beginning of acceleration phases, the temperature curve is smoother. While accelerating out of corners, a further advantage can be gained if a controller is used simultaneously. Using a higher integrating and no differentiating part for the PID controller, a certain temperature overshoot can be realized, which results, with regard to a mechanical water pump, in a slightly higher heat saving. According to the lumped capacity model, a heat saving potential about one kilowatt is possible. This corresponds to a share of approximately 2.5 percent of the total heat loss.

\subsection{Temperature shifting and resulting heat saving}
\label{ShiftingEffect}

\begin{figure}[H]
\captionsetup{width=1.0\textwidth}
\begin{center}
	 \includegraphics[width=1.0\textwidth]{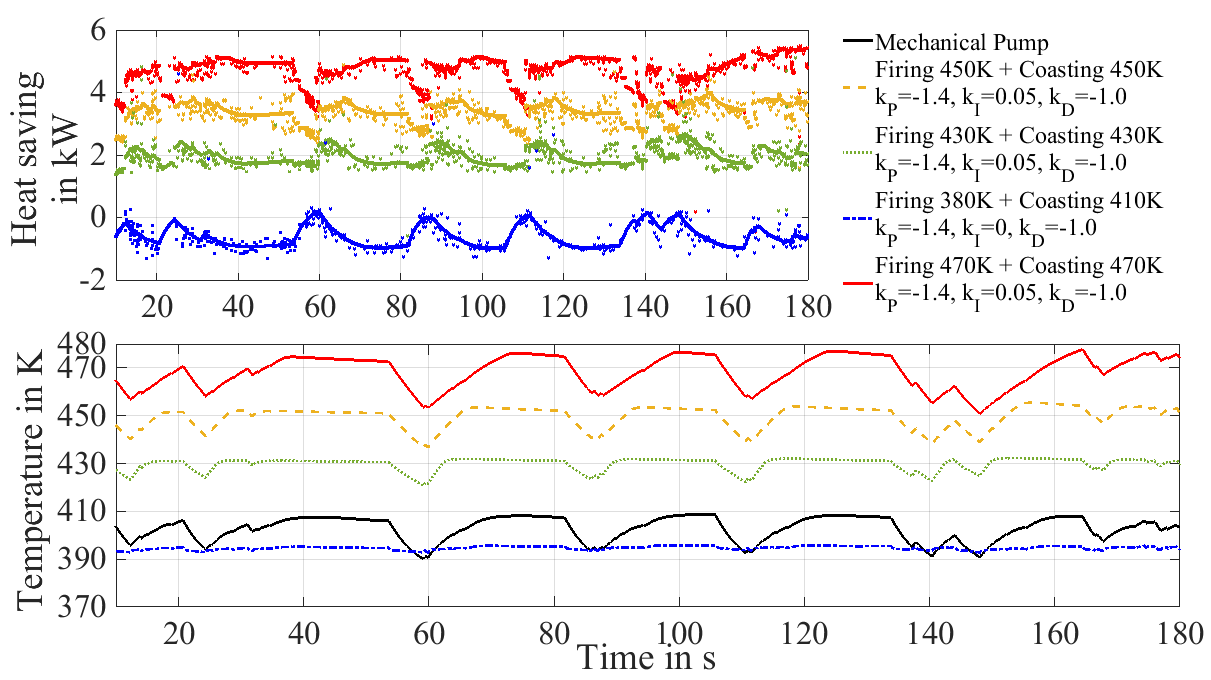}
\end{center}
\caption{Lumped capacity model: comparison of various control strategies and target temperatures concerning heat saving potentials and temperature consistency under a transient drive.}
\label{fig:DynamicII_Heat}
\end{figure}

Next, different target temperatures are used and compared to a mechanical water pump. As can be seen in fig. \ref{fig:DynamicII_Heat}, according to the lumped capacity model, a heat saving potential of five kilowatts can be achieved by shifting the temperature up to 470 K. This corresponds to a share of approximately 10 percent of the total heat loss. Interestingly, a very smooth temperature curve can be realized if one uses anticyclical target temperatures, e.g., a higher target temperature under coasting conditions than under fired situations.

\subsection{Three-dimensional results}
\label{Threedimen}

In the following, according to fig. \ref{fig:Temperature_Method}, transient finite volume simulations are presented compared with the lumped capacity model.

\begin{figure}[H]
\captionsetup{width=1.0\textwidth}
\begin{center}
	 a) \includegraphics[width=0.45\textwidth]{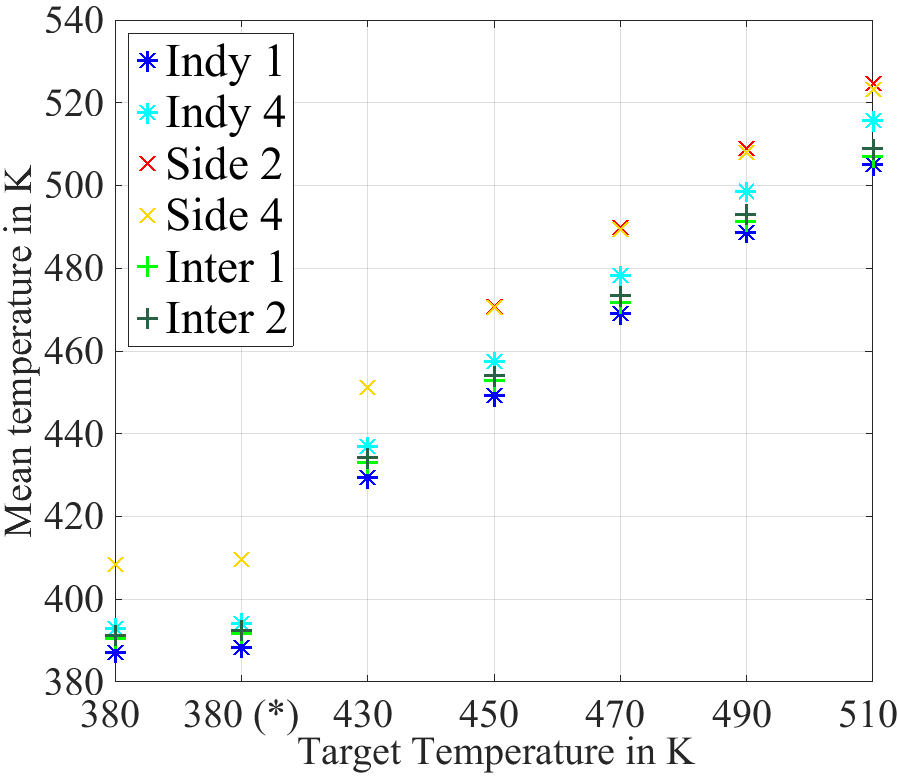} 
	 b) \includegraphics[width=0.45\textwidth]{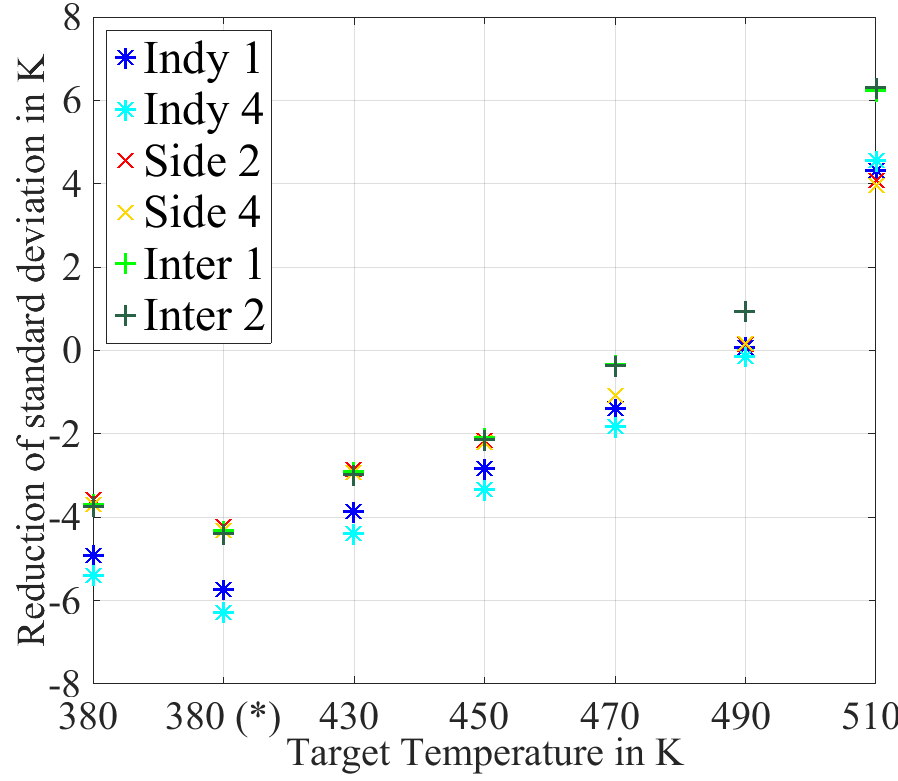}
\end{center}
\caption{Detailed 3D simulation: feasible temperature range by using a PID controller with $k_{\text{P}}=-1.4~\text{kg/(Ks)}$, $k_{\text{I}}=-0.05~\text{kg/(Ks\textsuperscript{2})}$ and $k_{\text{D}}=-1.0~\text{kg/K}$. a) Mean temperature. b) Mean differences in the temperature standard deviation with regard to a mechanical pump. The ratio of the (reference) mechanical water pump was set in order to reach a maximum temperature of 407 K.}
\label{fig:FeasibleRange3D}
\end{figure}

\begin{figure}[H]
\captionsetup{width=1.0\textwidth}
\begin{center}
	 \includegraphics[width=0.8\textwidth]{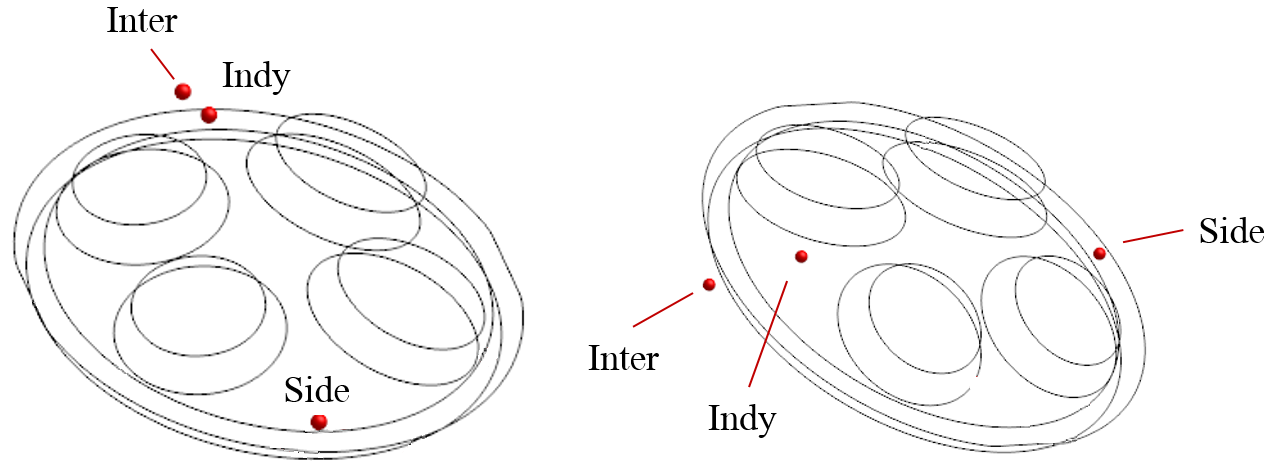} 
\end{center}
\caption{Overview of the measurement positions in the detailed FVM simulation.}
\label{fig:Overview_MeasurementPositions}
\end{figure}

Compared to fig. \ref{fig:FeasibleRange}, mean temperature results, together with their standard deviations, for the detailed FVM simulation are shown in fig. \ref{fig:FeasibleRange3D}. In the case of the detailed simulation, one can notice a larger temperature range for the feasible mean temperature. One possible reason can be the additional heat flux due to heat conduction. It seems that, with higher mean temperatures, its relative weighting falls in relation to the heat flux due to water cooling. This can be explained with the higher overall temperature level. As a consequence, the water cooling becomes more important, e.g., the temperature difference between solid and water has to be larger. With regard to fig. \ref{fig:FeasibleRange} b) with fig. \ref{fig:FeasibleRange3D} b), one can notice a smaller sensitivity with regard to the temperature variance. One argument can also be the additional heat flux due to heat conduction, and, therefore, the lower sensitivity towards water mass flow rates. In addition, the temperature gradients in the detailed simulation cause a kind of filtering function. In the case of the lumped model, there exist no spatial gradients, and, therefore, every change in the boundary conditions directly effects the solid temperature.
An overview of the used measurement positions is given in fig. \ref{fig:Overview_MeasurementPositions}: all positions are located within the cylinder head and one millimeter below the surfaces. The numbers indicate different cylinders. In this case, the command variable was related to the measurement position "Indy 1". The second setting which is called 380 K (*) corresponds to the anticyclical target temperature strategy from fig. \ref{fig:DynamicII_Heat}. For the detailed simulation, it can be observed that a larger feasible temperature range can be realized. Temperature values up to 510 K are possible. Compared to the mechanical water pump, lower values for the standard deviation of the engine temperatures are reachable. On the other hand, the increase with higher target temperatures is not that high. A key element appears to be that the other five measurement points near the combustion chamber behave in a manner similar to that of the command variable "Indy 1".

\begin{figure}[H]
\captionsetup{width=1.0\textwidth}
\begin{center}
	 a) \includegraphics[width=0.45\textwidth]{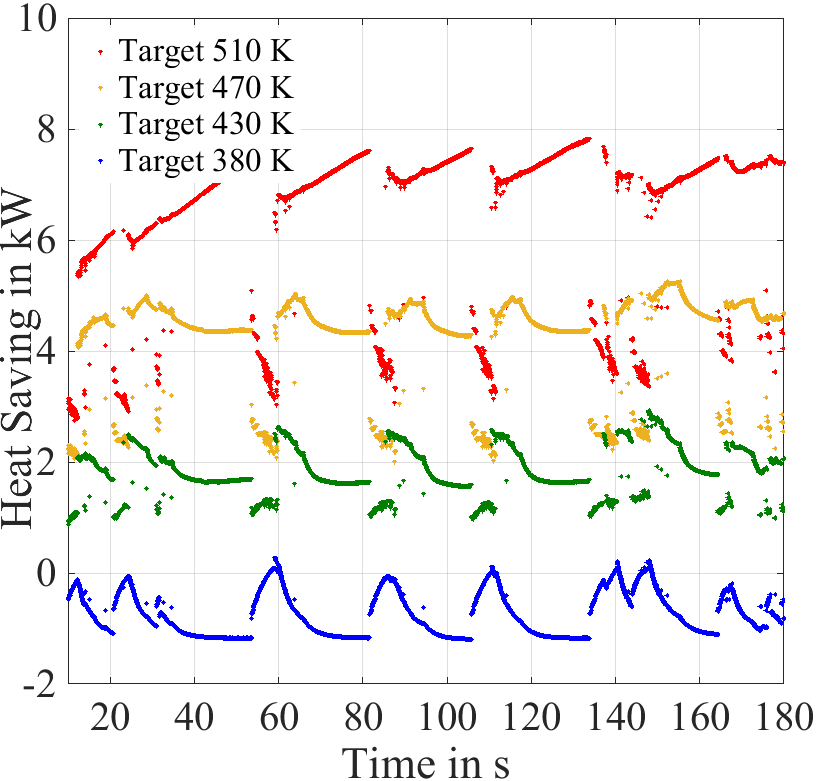} 
	 b) \includegraphics[width=0.45\textwidth]{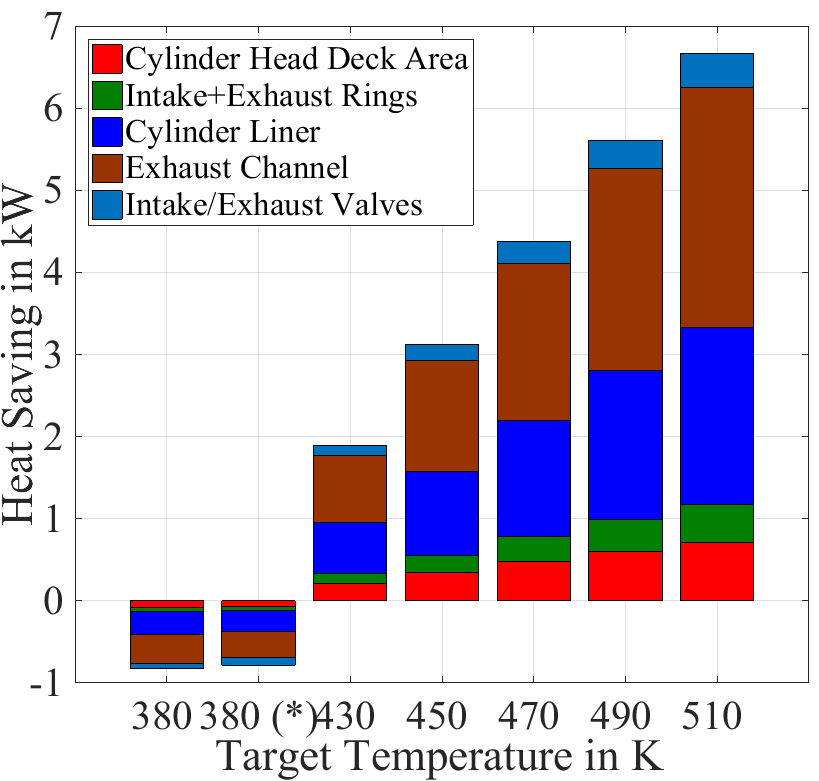}
\end{center}
\caption{Detailed 3D simulation: a) Heat saving potential with regard to a mechanical pump. The PID controller parameters are the same as in fig. \ref{fig:FeasibleRange3D}. a) Transient heat saving potential. b) Mean heat saving potential with corresponding spatial distribution. The ratio of the (reference) mechanical water pump was set in order to reach a maximum temperature of 407 K.}
\label{fig:3D_Shifting}
\end{figure}

Compared with fig. \ref{fig:DynamicII_Heat}, transient heat saving potentials for the detailed simulations are shown in fig. \ref{fig:3D_Shifting}. Under dynamic operating conditions, heat savings up to 8 kW are possible for a target temperature of 510 K. This corresponds to a share of approximately 15 percent of the total heat loss. Because of the large standard deviation at such high temperature levels, this potential drops below 4 kW under coasting conditions. 

As can be seen in fig. \ref{fig:3D_Shifting} b), most of the heat is saved in the exhaust channels and the cylinder liners. Surprisingly, the cylinder head deck area has a small proportion. The higher heat saving potential in the case of the detailed simulation, as can be seen in fig. \ref{fig:DynamicII_Heat} and fig. \ref{fig:3D_Shifting}, is a result of higher, feasible mean temperatures.

\begin{figure}[H]
\captionsetup{width=1.0\textwidth}
\begin{center}
	 \includegraphics[width=1.0\textwidth]{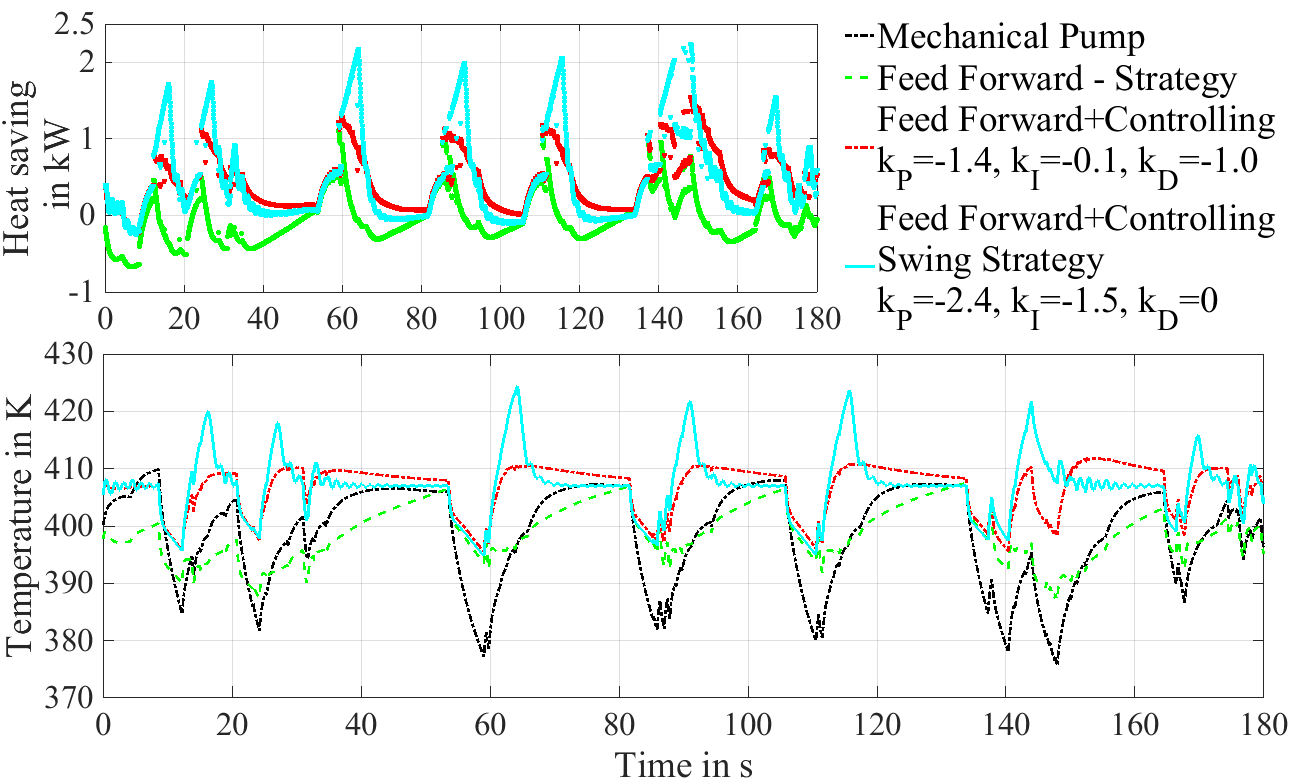} 
\end{center}
\caption{Detailed 3D simulation: comparison of four different feed forward and control strategies concerning heat saving potentials under a transient drive.}
\label{fig:DynamicI3D}
\end{figure}

Compared to fig. \ref{fig:DynamicI}, the detailed FVM simulations show a higher transient heat saving potential in acceleration phases. As can be seen in fig. \ref{fig:DynamicI3D}, the temperature swing strategy delivers values up to 2 kW. Regarding the integrating part of the controller, it is important to note, however, that the oscillation behavior is more pronounced.

Interestingly, with respect to the lumped capacity model, the difference between the pure feed forward strategy and the combination of pre-control and feedback control is higher. In general, the lumped capacity model underestimates the temperature swings: the lower temperatures during coasting are overestimated in the lumped model. One possible explanation can be the missing heat flux due to heat conduction, which is only contained in the three-dimensional simulation. As a result of this lower temperature difference between the mechanical pump and the control strategy, the heat saving potential is higher for the detailed simulation. In addition, the real, inhomogeneous distribution of the heat flux vector around the combustion chamber cannot be realized with a lumped model approach.

\section{Conclusions}
\label{conc}
\setlength{\parindent}{0pt}
\setlength{\parskip}{0pt}

Concerning the initial research questions, the results are as follows.
Within the limits of 0.25 and 4.5 kg/s for the water mass flow rate, a feasible range for the mean temperatures of engine components near the combustion chamber is 130 K. Compared with a conventional mechanical water pump, the standard deviation can be reduced up to 6 K, which can reduce possible damage mechanisms through local plastic deformations, e.g., low cycle fatigue or thermo-mechanical fatigue, respectively. For the highest mean temperature, the standard deviation was 6 K higher. A very important finding was the fact that, by using one command variable, the other measurement points near the combustion chamber behave in a similar manner. Concerning heat fluxes in full load states, it could be shown that a heat saving potential about 8 kW is possible. This corresponds to a share of approximately 15 percent of the total heat loss. In re-acceleration phases, values about 2 kW are realistic. Most of the heat losses can be reduced in the exhaust channels and the cylinder liners:  75 percent of the complete heat flow is reduced in this area. Of the remaining 25 percent, half of the heat is reduced at the cylinder head deck area. Valves and their rings contribute equally to the rest.\\
Concerning the two different modelling ideas, the qualitative progressions of relevant physical quantities are very similar. Therefore, the lumped capacity model can be used for optimization and sensitivity calculations: the Monte Carlo simulation has shown valuable correlations for the control device parameters, which could be effectively used in subsequent three-dimensional simulations. Compared to these detailed simulations, the feasible range for the mean temperatures is underestimated by 50 K. Analogously, the profit concerning standard deviations is slightly underestimated by 1 K. However, for high temperature levels, it is overestimated by 6 K. With regard to heat saving potentials, the simplified model gives too low values for both the dynamic and the quasi-stationary case: In the first one, the difference is 3 kW, in the second one, it is 1 kW.\\
The direct influence on the combustion, either positively affected by reduced heat losses, and stabilized temperature conditions, or negatively affected by higher inlet air temperatures, resulting in different knocking tendencies and volumetric efficiencies, has to be investigated experimentally. In addition, the model validity and its scope of application, shown in fig. \ref{fig:Parameter_Space_Water}, should be extended by the laminar transition regime, as well as by higher frequencies.

\section{Nomenclature}
\begin{longtable}{l p{0.6\textwidth} r}
	\hline
	\textbf{Symbol} & \textbf{Description} & \textbf{Unit} \\
	\hline
	\endhead
	\\
	 $A$ & Realisation of random variable $\alpha$ & W/(m\textsuperscript{2}K)\\
	 \\	
	 $C_{\text{v}}$ & Specific heat at constant volume & J/(kgK)\\
	 \\	
	 $C_{\text{p,w}}$ & Specific heat at constant pressure for water & J/(kgK)\\
	 \\	
	 $f$ & Frequency & Hz\\
	 \\	
	 $G_{\text{p}}$ & Transfer function for the water pump & dimensionless\\
	 \\
	 $G_{\text{c}}$ & Transfer function for the controller & kg/(Ks)\\
	 \\
	 $G_{\text{o}}$ & Transfer function for the open control loop & kg/(Ks)\\
	 \\
	 $G_{\text{cl}}$ & Transfer function for the closed control loop & kg/(Ks)\\
	 \\ 
	 $k_{\text{P}}$ & P-component of the controller & kg/(Ks)\\
	 \\	
	 $k_{\text{I}}$ & I-component of the controller & kg/(Ks\textsuperscript{2})\\
	 \\	
	 $k_{\text{D}}$ & D-component of the controller & kg/K\\
	 \\		
	 $k_{\text{s,e}}$ & Static amplification engine & (Ks)/kg\\
	 \\		
	 $m$ & Reynolds exponent & dimensionless\\
	 \\	
	 $\dot{m}_{\text{w}}$ & Water mass flow rate & kg/s\\
	 \\	
	 $\dot{m}_{\text{n,w}}$ & Nominal value of the water mass flow rate & kg/s\\
	 \\		  
	 $\dot{m}_\text{w}|_{\text{max}}$ & Maximum water mass flow rate & kg/s\\
	 \\	
	 $\dot{m}_\text{w}|_{\text{min}}$ & Minimum water mass flow rate & kg/s\\	 
	 \\		 
	 $\dot{m}_{\text{w}}|_{\text{ref}}$ & Reference water mass flow rate & kg/s\\
	 \\	 	 
	 $n$ & Engine speed & rounds per minute [rpm] \\
	 \\	
     $p_{\alpha | n}$ & Conditional probability density function on $\alpha$ with regard to $n$ & dimensionless\\	 
     \\     
	 $Pr$ & Prandtl number & dimensionless\\
	 \\		
	 $\dot{Q}$ & Heat flux & W\\
	 \\	
	 $Re$ & Reynolds number & dimensionless\\
	 \\	
	 $s$ & Laplace coordinate & 1/s\\
	 \\	
     $t$ & Physical time & s\\
	 \\
	 $T_{\text{Cyl}}$ & Cylinder head temperature & K\\
	 \\	
	 $T_{\text{Cyl,t}}$ & Command variable of the cylinder head temperature & K\\
	 \\		 
	 $T_{\text{gas}}$ & Reference (gas) temperature & K\\
	 \\	 
	 $T_{\text{mod}}$ & Statistically modified reference (gas) temperature & K\\
	 \\		
	 $T_{\text{w,i}}$ & Inlet water temperature & K\\
	 \\ 	 
	 $T_{\text{w}}$ & Water reference temperature & K\\
	 \\ 
	 $T_{\text{R,1}}$ & Time constant of control device & s\\
	 \\ 
	 $T_{\text{R,2}}$ & Time constant of control device & s\\
	 \\  
	 $Y$ & Laplace transform of $y$ & various\\
	 \\
	 $y$ & Arbitrary function & various\\
	 \\ 	
	 $z$ & Disturbances of all kinds on the engine & various\\
	 \\ 	  
	 $\fx$ & Position vector & m\\
	 \\   	 	
	 $\Delta x$ & Characteristic wall thickness & m\\
	 \\   		  	 	 	
\end{longtable}

\section*{Greek symbols}
\begin{longtable}{l p{0.6\textwidth} r}
	\hline
	\textbf{Symbol} & \textbf{Description} & \textbf{Unit} \\
	\hline
	\endhead
	\\
	 $\alpha$ & Heat transfer coefficient & W/(m\textsuperscript{2}K)\\
	 \\
	 $\alpha_{\text{ref}}$ & Reference heat transfer coefficient & W/(m\textsuperscript{2}K)\\
	 \\	 
	 $\alpha_{\text{w}}$ & Effective heat transfer coefficient for the water channel & W/(m\textsuperscript{2}K)\\	 
	 \\	 
	 $\alpha_{\text{c}}  $ & Effective heat transfer coefficient for the combustion chamber & W/(m\textsuperscript{2}K)\\	 
	 \\	 	 
	 $\rho$ & Mass density & kg/m\textsuperscript{3}\\
	 \\	
	 $\tau_{\text{p}}$ & Time constant of water pump & s\\
	 \\	 
	 $\tau_{\text{e}}$ & Time constant of engine & s\\
	 \\	 
	 $\tau_{\text{c}}$ & Time constant of closed loop & s\\
	 \\	 	 
	 $\chi$ & Ratio between effective surfaces of the combustion chamber and the water jacket & dimensionless\\  
\end{longtable}

\section*{Mathematical Notation}
\begin{longtable}{p{0.6\textwidth} p{0.0\textwidth}  r}
	\hline
	\textbf{Symbol} & \textbf{} & \textbf{Description} \\
	\hline
	\endhead
	\\	
	 $\pd{\left(\cdot\right)}{t}$ & & Partial time derivative  \\
	 \\	 	 
	 $\pd{\left(\cdot\right)}{\fx}$ & & Spatial gradient \\
	 \\	 	 
	 $\langle{\cdot}\rangle$ & & Expectation value regarding time  \\
\end{longtable}

\section*{Abbreviations}

\begin{longtable}[c]{p{0.35\textwidth} p{0.0\textwidth}  r}
	\hline
	\textbf{Abbreviation} & \textbf{} & \textbf{Description} \\
	\hline
	\endhead
	\\
	$ CFD    $ & & \textbf{C}omputational \textbf{F}luid \textbf{D}ynamics\\	
	$ CHT $ & & \textbf{C}onjugate \textbf{H}eat \textbf{T}ransfer\\	
	$ FVM $ & & \textbf{F}inite \textbf{V}olume \textbf{M}ethod  \\
	$ HTC $ & & \textbf{H}eat \textbf{T}ransfer \textbf{C}oefficient \\
	$ ACT $ & & \textbf{A}verage \textbf{C}ylinder \textbf{T}emperature \\
\end{longtable}

\section{References}
\bibliography{library}
\end{document}